\documentclass[12pt,a4paper,final]{iopart}
\pdfoutput=1
\def\contentsname{\empty}
\usepackage{graphicx}
\usepackage{cite}
\usepackage{bigfoot}
\usepackage{appendix}
\usepackage[breaklinks=true,colorlinks=true,linkcolor=black,urlcolor=black,citecolor=black,pdfencoding=auto]{hyperref}
\usepackage{amsfonts, amsmath, dsfont}
\usepackage{empheq}

\usepackage{bookmark}% faster updated bookmarks

\def\contentsname{\empty}

\newcommand{\bea}{\begin{eqnarray}}
\newcommand{\eea}{\end{eqnarray}}

\newcommand{\be}{\begin{equation}}
\newcommand{\ee}{\end{equation}}

\def\Ai{\text{Ai}}

\newcommand{\nn}{{\nonumber}}
\def\be{\begin{equation}}
\def\ee{\end{equation}}

\makeatletter
\def\@fnsymbol#1{\ifcase#1\or \dagger\or \ddagger\or \S\or
   \|\or \P\or ^{+}\or ^{\tsty *}\or \sharp
   \or \dagger\dagger \or \mbox{$\clubsuit$} \else\@ctrerr\fi\relax}
\makeatother

\begin{document}

%\title[Overlaps of the N{\'e}el state with XXZ Bethe states]{Generalized Gaudin determinant for the overlap of the N{\'e}el state with XXZ Bethe states}
%\title[Dynamical correlations of the Directed polymer]{Dynamical correlations of the Directed polymer }
\begin{center}
\title[Two-time height distribution
for 1D KPZ growth]{Two-time height distribution for 1D KPZ growth: \\
the recent exact result and its tail via replica}
\end{center}
%\title[Preparing an article for IOP journals in  \LaTeXe]{Working title: Overlaps with the N{\'e}el state and gTBA}

\author{Jacopo de Nardis$^1$ and Pierre Le Doussal$^2$}
\address{$^1$ D\'epartement de Physique, Ecole Normale Sup\'erieure, PSL Research University, CNRS, 24 rue Lhomond, 75005 Paris, France \\
$^2$ CNRS-Laboratoire de Physique Th\'eorique de l'Ecole Normale Sup\'erieure, 24 rue Lhomond, 75231 Paris Cedex, France}
\ead{jacopo.de.nardis@phys.ens.fr}
\ead{ledou@lpt.ens.fr}

\begin{abstract} 
We consider the fluctuations in the stochastic growth of a one-dimensional interface of height $h(x,t)$ described by the Kardar-Parisi-Zhang (KPZ)
universality class. We study the joint probability distribution function (JPDF) of the interface heights
at two-times $t_1$ and $t_2>t_1$, with droplet initial conditions at $t=0$.
In the limit of large times this JPDF is expected to become a universal function of the time ratio $t_2/t_1$, and of the (properly scaled) heights $h(x,t_1)$ and $h(x,t_2)$. Using the replica Bethe ansatz method for the KPZ equation,
in \cite{two-time-long-us} we obtained a formula for the JPDF in the (partial) tail regime where $h(x,t_1)$ is large and positive, subsequently found in excellent agreement with experimental and numerical data \cite{two-time-letter-us}. Here we show that our results are in perfect agreement with Johansson's recent rigorous expression of the full JPDF \cite{Johansson2times-final}, thereby confirming the validity of our methods.

\end{abstract}

% %Uncomment for PACS numbers title message
% \pacs{00.00, 20.00, 42.10}
% % Keywords required only for MST, PB, PMB, PM, JOA, JOB?
% \vspace{2pc}
% \noindent{\it Keywords}: Article preparation, IOP journals
% % Uncomment for Submitted to journal title message
% \submitto{\JPA}
% % Comment out if separate title page not required
% %\maketitle

%TODO:
%- Pierre maybe wants to change formula \eqref{explicit}
%- Add persistence story in the abstract
%- Misprint below Eq 138
%"once inserted in equation E.5" JACOPO: corrected.

%\tableofcontents

\section{Introduction}

Growth processes are natural phenomena occurring when a stable phase of a generic system expands into a non-stable (or meta-stable) one, generally in presence of noise. Such interface separating the two phases develops many non-trivial geometric and statistical features during its time evolution. In two dimension a universal macroscopic 
behavior emerges, 
unifying many growth phenomena into a few universality classes, irrespective of their microscopic details. The most generic one, for stochastic growth rules,
is the celebrated Kardar-Parisi-Zhang universality class. A prominent member of this
class is the continuum KPZ equation \cite{KPZ}
\begin{equation} \label{kpzeq0} 
\partial_t h(x,t) = \nu \partial_x^2 h(x,t) + \frac{\lambda_0}{2}  (\partial_x h(x,t))^2 + \sqrt{D} ~ \eta(x,t) \ ,\nonumber
\end{equation}
which describes the motion of an interface of height $h(x,t)$ at point $x \in \mathbb{R}$
%as a function of 
at time $t$, %and 
driven by a unit space-time white noise $\eta(x,t)$. In the past years many tools have been developed to solve this equation. Its solution is described at large times as a uniformly moving front with sub-leading 
$O(t^{1/3})$ fluctuations, as $h(x,t) = v  \,t + t^{\frac{1}{3}} \tilde{h}(x,t) + O(1)$, where $v$ is a parameter of the model and $\tilde{h}(x,t)$, the scaled height field, is a stochastic variable whose statistical properties become universal at large $t$. Numerous exact results were found \cite{spohn2000,we,dotsenko,spohnKPZEdge,corwinDP,sineG,we-flat,we-flatlong,cl-14,dotsenkoGOE,Quastelflat,SasamotoStationary,BCFV,Cor12,KPZFixedPoint2}, as well as experimental and numerical results \cite{exp4,Takeuchi,TakeuchiHHLReview}, showing the convergence
at large time of the {\it one-time} probability distribution function (PDF) of $\tilde{h}(x,t)$ towards
a few universal distributions, the Tracy Widom (TW) distributions which also appear in random matrix theory.
One approach is to map the solution of the KPZ equation to the free energy of a directed polymer (DP) in a random potential and then use standard replica methods to compute the integer moments, $\overline{Z^n}$, of the DP partition sum \cite{we,sineG,we-flat,we-flatlong,cl-14,dotsenkoGOE}. From these integer moments
one extracts the probability distribution function (PDF) of the KPZ height field, a 
non-rigorous but generally correct approach. The average over the random potential leads to an effective interaction among the replicas, described by the attractive Lieb-Liniger (LL) Hamiltonian \cite{ll}. Therefore computing $\overline{Z^n}$ reduces to computing quantum transition amplitudes in imaginary time in the attractive LL model. Since this model can be solved via the Bethe ansatz, the method is called the {\it replica Bethe ansatz} (RBA). The ground state of the LL model is a single bound state containing all the particles (replicas), while excited states are obtained by splitting this bound state in several smaller ones \cite{cc-07}. To perform the calculation one expands over the complete set of 
eigenstates and re-sum the series. The contribution of the ground state is not sufficient to 
reproduce the long time limit of the KPZ equation (despite what intuition may suggest), and 
the other excited states need to be included in order to obtain the full statistics 
of the fluctuations at large times. This summation is not easy, but could be performed
for one-time observables. For two-time observables however, such as the joint probability distribution (JPDF) of the scaled heights at two different times, $\tilde h(0,t_1)$ and $\tilde h(0,t_2)$, the calculation is very difficult
\cite{dotsenko2times1,dotsenko2times2,dotsenko2times3,Johansson2times,FerrariSpohn2times,BaikTASEP}, and 
almost no practically useful results (i.e. which could be numerically evaluated and compared to numerical simulations and experiments)
existed until recently.

In a recent work \cite{two-time-long-us,two-time-letter-us}
we obtained a formula for the tail of the two-time JPDF,
for the droplet, or wedge, initial condition $h(x,t=0) = - w |x|$ in the large time limit.
We started from the observation that the right tail of the {\it one-time} PDF of $\tilde{h}(x,t)$ 
at large $t$, i.e. of the TW distribution, can be well approximated from a proper treatment of the
contribution of the LL ground state. We then conjectured that a similar property holds also for the
two-time JPDF. That allowed us to perform a partial summation over states and 
obtain the following result. Consider the limit where both times $t_1$, $t_2$ are large 
but their ratio is kept finite $t_2/t_1 = 1+ \Delta$. In that limit the joint cumulative probability
distribution function (JCDF) becomes a function of $\Delta$ only 
\bea  
&&  \lim_{t_1 \to \infty} \text{Prob}(\tilde{h}(x_1,t_1) < \sigma_1, \tilde{h}(x_2,t_1 (1+\Delta)) < \sigma_2) 
 =\hat{\mathcal{P}}_{\Delta}(\sigma_1, \sigma_2) ,\\
&& \hat{\mathcal{P}}_{\Delta}(\sigma_1, \sigma_2) = \mathcal{P}_{\Delta}(\sigma_1, \sigma) 
\quad , \quad \sigma_2=\frac{\sigma_1+\sigma \Delta^{1/3}}{(1+ \Delta)^{1/3}} ,
\eea
which we found more convenient to express (see second line) as a 
function of $\sigma$, which is associated to the scaled height difference (see below),
a simple linear combination of $\sigma_1$ and $\sigma_2$. Our result
was that for large positive $\sigma_1$ and any $\sigma$, $\Delta$
\begin{equation} \label{tailapp2} 
\mathcal{P}_{\Delta}(\sigma_1, \sigma) =  \mathcal{P}^{(1)}_{\Delta}(\sigma_1, \sigma) + O(e^{- \frac{8}{3} \sigma_1^{3/2}})  ,
\end{equation}
and we obtained an explicit formula for $\mathcal{P}^{(1)}_{\Delta}(\sigma_1, \sigma)$,
see below.

The distribution $\mathcal{P}^{(1)}_{\Delta}$ was used to compute the conditional two-time covariance ratio
\begin{equation}
C_{\Delta,\sigma_c} = \lim_{t_1 \to \infty} \frac{\langle \tilde{h}(x,t_1)  \tilde{h}(x,t_1(1+\Delta)) \rangle^c_{\tilde{h}(x,t_1) > \sigma_c}}{\langle \tilde{h}(x,t_1)^2 \rangle^c_{\tilde{h}(x,t_1) > \sigma_c}},
\end{equation}
and to compare it with experimental and numerical results in \cite{two-time-long-us,two-time-letter-us}.
We found excellent agreement for values of $\sigma_c > -1$, i.e. well beyond the naive
range of validity, while the tail approximation starts to break down for smaller values
of $\sigma_c$ (we recall that $\sigma_c=-1.7...$ is the mean of the TW distribution). 
This quantity shows how the system is aging during its time evolution and it is of theoretical \cite{Johansson2times,FerrariSpohn2times,dotsenko2times1,LeDoussal-LargeDelta} and experimental interest \cite{TakeuchiCrossover,TakeuchiPersistence}. In particular its limit  $\Delta \to \infty$ and $\sigma_c \to -\infty$ quantifies the so-called ergodicity breaking of the KPZ class \cite{TakeuchiPersistence}. 
In a separate RBA calculation, extending a framework based on Airy processes also discussed
in \cite{two-time-long-us,FerrariSpohn2times} and valid for $\Delta \to +\infty$, it was obtained that for droplet initial conditions
the full two-time covariance ratio does not decay to zero in the large $\Delta$ limit, but instead 
to a universal number \cite{LeDoussal-LargeDelta}
\begin{equation}
\lim_{\sigma_c \to  - \infty} \lim_{\Delta \to \infty} C_{\Delta,\sigma_c} = 0.623 \ldots,
\end{equation}
in good agreement with previous experimental observations \cite{Y}, and in excellent
agreement with simultaneous numerical simulations \cite{X}. The calculation 
of \cite{LeDoussal-LargeDelta} also confirmed the result of \cite{two-time-long-us,two-time-letter-us} 
but only in the regime where both $\sigma_1$ and $\Delta$ are simultaneously large
(see also \cite{JohanssonNotes}). 

Despite all the numerical and experimental evidence, a rigorous proof of the tail $\mathcal{P}^{(1)}_{\Delta}$ provided in\cite{two-time-long-us,two-time-letter-us} for arbitrary $\Delta$ was still lacking up to now. Recently an exact expression for the full two-time cumulative distribution $\hat{\mathcal{P}}_{\Delta}(\sigma_1, \sigma_2)$ was obtained by Johansson in \cite{Johansson2times-final}. It is obtained from a scaling
limit of a calculation for a discrete DP model (last passage percolation), simpler
than the one in \cite{Johansson2times}. It leads to a much more tractable expression
than the one in \cite{Johansson2times}, in fact a (relatively) simple one, as a contour integral of 
a matrix Fredholm determinant. Hence, we are now in position to check our expression for the tail, in the limit where $\sigma_1$ is large and positive.  \\

In this paper we show that, by expanding the exact full two-time distribution of Ref. 
\cite{Johansson2times-final} at large and positive $\sigma_1$, the first order term of the expansion reproduces our result for the tail \cite{two-time-long-us,two-time-letter-us}. Our result therefore provides a non-trivial check of our expression for the tail and of the newly derived distribution \cite{Johansson2times-final}.

\section{KPZ equation and results for the one-time statistics of the height}
\label{sec:review}

While the large time results discussed here are expected to be valid for
the whole KPZ class, independently of the details of
the model, the explicit calculation performed in \cite{two-time-long-us,two-time-letter-us}
starts from the continuum KPZ equation \eqref{kpzeq0}. Using the units of space $x^* = \frac{(2 \nu)^3}{D \lambda_0^2}$, time $t^* = \frac{ 2 (2 \nu)^5}{D^2 \lambda_0^4}$ and height $h^* = \frac{2 \nu}{\lambda_0}$ it becomes 
\begin{equation} \label{kpzeq} 
\partial_t h(x,t) = \partial_x^2 h(x,t) +   (\partial_x h(x,t))^2 + \sqrt{2} ~ \eta(x,t)  ,
\end{equation}
with white noise $\overline{\eta(x,t) \eta(x',t')}=\delta(x-x') \delta(t-t')$. For the purpose
of the calculation, the droplet IC was implemented using the so-called hard-wedge initial condition defined as
$h_{w_0}(x,0) =  - w_0  |x| + \ln(\frac{w_0}{2})$ in the limit $w_0 \to +\infty$, so that $\exp(h_{w_0}(x,0)) \to \delta(x)$. Note that in the large time limit, the same universal results are expected for any finite $w_0>0$
or any other initial condition in the droplet class. 
At large time the KPZ field grows linearly in time with $O(t^{1/3})$ fluctuations.
The fluctuations at one space point (choosing here $x=0$) are
governed by the Tracy Widom distribution associated to the Gaussian unitary ensemble (GUE).
Its PDF is $f_2(\sigma) = F_2'(\sigma)$, i.e. one has at large $t$ \cite{spohn2000,we}
\bea
h(x=0,t) = - \frac{t}{12} + t^{1/3} \tilde{h} + o(t^{1/3}) \quad , \quad {\rm Prob}(\tilde{h} < \sigma) = F_2(\sigma),
\eea
where $F_2(\sigma)$ is given by the Fredholm determinant
\be \label{airyK1}
F_2(\sigma) = {\rm Det}[I - P_\sigma K_{\rm Ai} P_\sigma],
\ee
involving the Airy Kernel $K_{\Ai}$:
\be \label{airyK2}
  K_{\rm \Ai}(v,v')= \int_{0}^{+\infty} dy \Ai(y+v) \Ai(y+v')
= \frac{\Ai(v) \Ai'(v')-\Ai'(v) \Ai(v')}{v-v'},
\ee
where $P_\sigma(v)=\theta(v-\sigma)$ is the projector on $[\sigma,+\infty[$.\\

In \cite{two-time-long-us} we introduced the tail approximation of the CDF of the GUE-TW distribution as the function % \eqref{eq:gueapprox_eq}
\begin{equation} \label{TWtail}
F^{(1)}_2(\sigma)  \equiv  1 - \text{\text{Tr}} \, P_\sigma K_{\Ai} = 1 - \int_\sigma^{+\infty} dv K_{\rm Ai}(v,v),
\end{equation}
which keeps only the first term in the series expansion of the Fredholm determinant, 
hence it captures the leading (stretched) exponential behavior for large and positive $\sigma$,  $F^{(1)}_2(\sigma) -1= O(e^{- \frac{4}{3} \sigma^{3/2}})$ and the corrections are terms $O(K_{\rm Ai}^2)$
containing products of four Airy functions and more, of higher (stretched) exponential order
\be
F_2(\sigma) = F^{(1)}_2(\sigma)  + O(e^{- \frac{8}{3} \sigma^{3/2}})
\ee
which is the analog of \eqref{tailapp2} for the one-time distribution. 
As can be seen in Fig. 1 of  \cite{two-time-long-us} this approximation is very good (with error less than $10^{-3}$) for any $\sigma > -1$. This approximation also
consists in keeping only single string states in the RBA method, leading
to huge simplifications in the application of the method.

\section{Tail of the two-time joint distribution from replica Bethe ansatz} \label{sec_twotimeproblem}

\subsection{Definitions and scaled variables} \label{sec:def} 

Consider now the height at two different times $t=t_1$ and $t=t_2 > t_1$,
and two different space points $x=0$ and $x=X$, and denote
\be  \label{defH12}
 H_1 \equiv  h(0,t_1) - v t_1   \quad , \quad  H_2 \equiv  h(X,t_2)  - v t_2 \quad , \quad  H\equiv  H_{21} \equiv  H_2-H_1,
\ee
where $v=-1/12$ here. We have defined the difference of the two heights 
built over the time difference $t_{21}\equiv  t_2-t_1$. In the large time limit, when both $t_1$ and $t_2$ are sent to $+\infty$, the relevant parameter characterizing the JPDF of the two heights
will be the time difference rescaled by the earlier time, denoted as
\begin{equation}
\Delta = \frac{t_2 - t_1}{t_1} >0.
\end{equation}

Let us start with $X=0$, the general case being discussed below.
From the previous Section, we know that the two heights grow in time as $H_1 \sim t_1^{1/3}$ and $H_2 \sim t_2^{1/3}$, hence we introduce the rescaled heights $h_1$, $h_2$ and $h$, through
\bea \label{defh1}
\fl && ~~~~~~  H_1 = t_1^{1/3} h_1 %\equiv \lambda h_1
\quad , \quad H_2 = t_2^{1/3} h_2 \quad , \quad H_2  - H_1    = (t_2-t_1)^{1/3}  h  = \Delta^{1/3}  t_1^{1/3}  h . \label{defh}
\eea
Here $ h_1,  h_2,  h$ are random variables. We note that
\be
h_2 = \frac{h_1 + h \Delta^{1/3}}{(1+\Delta)^{1/3}} \quad \Leftrightarrow 
h =h_2 (1 + \frac{1}{\Delta})^{1/3} - h_1 \Delta^{-1/3}.
\ee 
From the previous section, the PDF of $h_1$ 
is the GUE-TW, $f_2$, and similarly for $h_2$. 
The JCDF of $h_1$ and $h_2$ in the large time limit was defined in the 
introduction as 
\be  \label{defP}
\hat {\cal P}_\Delta(\sigma_1,\sigma_2) =
\lim_{t_1,t_2 \to +\infty,\frac{t_1-t_2}{t_1}=\Delta} \text{Prob}(h_1 < \sigma_1,  h_2 < \sigma_2),
\ee
which depends only on $\Delta$. The variables which
appear naturally in our result (and in Johansson's result, see below),
are $\sigma_1$ and $\sigma=\sigma_2 (1 + \frac{1}{\Delta})^{1/3} - \sigma_1 \Delta^{-1/3}$,
hence it is useful to define the function ${\cal P}_\Delta(\sigma_1,\sigma)$
\be \label{PPhat}
{\cal P}_\Delta(\sigma_1,\sigma) = 
\hat {\cal P}_\Delta \Big(\sigma_1,\sigma_2=\frac{\sigma_1 + \sigma \Delta^{1/3}}{(1+\Delta)^{1/3}}\Big),
\ee
which is simply \eqref{defP} expressed in the variables $\sigma_1,\sigma$. 
Note that { ${\cal P}_\Delta$} is {\it not the JCDF associated to $h_1$ and $h$.}
However from it one can obtain the JPDF of $h_1$ and $h$ defined as
\be
\label{defPP}
P_\Delta(\sigma_1,\sigma) = \lim_{t_1,t_2 \to +\infty,\frac{t_1-t_2}{t_1} =\Delta} \overline{ \delta(h_1-\sigma_1)
\delta(h-\sigma) },
\ee 
via the relation
\be \label{eqdiff} 
P_\Delta(\sigma_1,\sigma) = ( \partial_{\sigma_1} \partial_{\sigma} - \Delta^{-1/3} \partial_{\sigma}^2) 
{\cal P}_\Delta(\sigma_1,\sigma) ,
\ee
whose derivation is a simple change of variable detailed in \ref{app:jpdf}.

\subsection{Tail of the two-time distribution}

The calculation of the full distribution \eqref{defP}, \eqref{PPhat}, \eqref{defPP} using the
RBA method poses a formidable challenge via the replica method (a valid attempt was also conducted in \cite{dotsenko2times1,dotsenko2times2,dotsenko2times3} but it turned out to be incorrect, see \cite{two-time-long-us}). The difficulty comes from the summation over two sets of eigenfunctions of the LL model, 
and the complicated expression for the so-called form factors \cite{Slavnov,piroli-cal}. In \cite{two-time-long-us,two-time-letter-us} using a partial summation, where one of the set is restricted to the single string states
of the Lieb-Liniger model, we were able to derive a relatively simple expression for the tail approximation $\mathcal{P}^{(1)}_{\Delta}(\sigma_1, \sigma)$ of the cumulative distribution (which satisfies \eqref{tailapp2})
\begin{equation} \label{eq:OURTAIL}
\mathcal{P}^{(1)} (\sigma_1,\sigma)  =  1 + (F_2^{(1)}(\sigma_2)-1) + \left(  {F_2(\sigma)}  \text{\text{Tr}} \left[   \Delta^{1/3}  P_\sigma K^{\Delta}_{\sigma_1}   P_\sigma   (I-P_{\sigma} K_{\Ai}P_{\sigma} )^{-1}   - P_{\sigma_1} K_\Ai  \right] \right),
\end{equation}
with \footnote{The CDF $\hat{\mathcal{P}}=\mathcal{P}$ is $g_\Delta$ in Eq. (75) in \cite{two-time-long-us},
where we corrected a misprint (immaterial for the main result Eq. (23) there), together with 
the tail approximation Eq. (139) there.} $\sigma_2 = \frac{\sigma_1 + \sigma \Delta^{1/3}}{(1+\Delta)^{1/3}}$,
expressed in terms of the Airy kernel, as well as of a novel kernel
\begin{equation} \label{K4def0}
K^{\Delta}_{\sigma_1}(u,v)   =  \int_0^\infty dy_1 dy_2 \Ai\left(-  {y_1}{ }   +  u \right) K_{\Ai}(y_1 \Delta^{1/3} + \sigma_1, y_2\Delta^{1/3} + \sigma_1)  \Ai \left(-  {y_2}{}  +  v \right),
\end{equation}
where we recall that $F_2(\sigma)$ and $F^{(1)}_2(\sigma_1)$ are respectively the GUE-TW CDF and its tail
approximation, given respectively
by (\ref{airyK1}) and (\ref{TWtail}).  

The corresponding tail approximation JPDF was displayed in \cite{two-time-long-us} Eq. 23
as 
\be
P^{(1)}_\Delta(\sigma_1,\sigma) = ( \partial_{\sigma_1} \partial_{\sigma} - \Delta^{-1/3} \partial_{\sigma}^2) 
{\cal P}^{(1)}_\Delta(\sigma_1,\sigma) .
\ee
Our result for the JPDF satisfies two important properties.
In the limit of infinite time difference $t_2/t_1 \to +\infty$ (which corresponds to the limit $\Delta \to + \infty$), it converges to the product of two GUE-TW
distributions
\begin{align} \label{prodGUE}
 \lim_{\Delta \to \infty} P^{(1)}_{\Delta}(\sigma_1,\sigma)   =  F_2^{(1) \prime }(\sigma_1)  \ F_2'(\sigma),
\end{align}
and in the limit of small (scaled) time separation $(t_2-t_1)/t_1 \ll 1$ (which translates into $\Delta \to 0^+$) it also decouples as follows
\begin{equation} \label{decoupleBR}
\lim_{ \Delta \to 0} P^{(1)}_{\Delta}(\sigma_1,\sigma)  =     F_2^{(1) \prime}(\sigma_1)    \   F_0'(\sigma),
\end{equation}
where $F_0(\sigma)$ is the Baik-Rains { (BR)} cumulative distribution \cite{png,SasamotoStationary} which
governs the stationary growth profile in the infinite time limit.\\

\section{Johansson's result for the full two-time  joint probability distribution}

Here we report Johansson's recent result \cite{Johansson2times-final} in our present notations. 
We focus here on the joint distribution for the two heights at the same position $X=0$.  There the following functions are defined 
\begin{equation}
S_1(x,y) = - \Delta^{-1/3} e^{(y-x)\delta}  \int_0^\infty  ds K_{\Ai}(\sigma_1 -s , \sigma_1 - x) K_{\Ai}(\sigma + s \Delta^{-1/3}, \sigma + y \Delta^{-1/3}),
\end{equation}
\begin{equation}
T_1(x,y) =  \Delta^{-1/3} e^{(y-x)\delta}  \int_{-\infty}^0  ds K_{\Ai}(\sigma_1 -s , \sigma_1 - x) K_{\Ai}(\sigma + s \Delta^{-1/3}, \sigma + y \Delta^{-1/3}),
\end{equation}
\begin{equation}
S_2(x,y) = \Delta^{-1/3} e^{\delta(y-x)}  K_{\Ai}(\sigma + x \Delta^{-1/3}, \sigma + y \Delta^{-1/3}),
\end{equation}
\begin{equation} \label{defS3}
S_3(x,y) = e^{\delta(y-x)} K_{\Ai}(\sigma_1 - x, \sigma_1 - y ),
\end{equation}
Notice that the regulator $\delta >0$ can be set to $0^+$. 
Moreover the following composite kernels are introduced
\bea \label{defST} 
&& S(x,y) = S_1(x,y)  + \theta(x) S_2(x,y) - \theta(-y) S_3(x,y), \\
&& T(x,y) = -T_1(x,y) - \theta(x) S_2(x,y) + \theta(-y) S_3(x,y).
\eea
The full kernel is then given by, defining $R_u(x,y) = S(x,y) + u^{-1} T(x,y)$
\begin{equation}\label{kernelKurt0}
K_u =  
\begin{pmatrix} P^- R_u P^-   &  P^- R_u P^+ \\
u P^+ R_u P^- & u P^+ R_u P^+
\end{pmatrix} ,
\end{equation}
where we introduced the projectors $P^{+}$ on the interval $[0,\infty)$ and $P^{-}$ on the interval $(-\infty,0]$. 
Then theorem 2.1 plus the formula (15) in \cite{Johansson2times-final} shows that the joint probability distribution at two different times can be written as the contour integral of a Fredholm determinant
\bea \label{eq:KURTRESULT}
  \hat {\cal P}_\Delta(\sigma_1,\sigma_2) &=& \lim_{t_1,t_2 \to +\infty,\frac{t_1-t_2}{t_1} =\Delta}
{\rm Prob}(h_1< \sigma_1, h_2< \sigma_2) \\
& =& \frac{1}{2 \pi i }\int_{R_1} \frac{d u}{ u-1} \text{Det}(1 + K_{u}) \\
& = & {\cal P}(\sigma_1,\sigma),
\eea
with $R_1$ a circle of radius larger than $1$. As expressed by the last line, this formula, quite remarkably,
is expressed naturally as a function of 
of $\sigma_1$ and $\sigma=\sigma_2 (1 + \frac{1}{\Delta})^{1/3} - \sigma_1 \Delta^{-1/3}$, a property
already found in our tail result, as discussed above.

Before embarking on calculations it is useful to define the following functions, defined
via a similarity transformation
\bea \label{similarity}
S_1(x,y)= \Delta^{-1/3} \tilde S_1(x \Delta^{-1/3}, y \Delta^{-1/3}),
\eea 
and the same definition (and tilde notation) for $S_2$ and $T_1$, leading to
\bea \label{STsim} 
&& \!\!\!\!\!\!\!\! \tilde S_1(x,y) = - \Delta^{1/3}  \int_0^\infty  ds K_{\Ai}(\sigma_1 -s \Delta^{1/3} , \sigma_1 - x \Delta^{1/3}) K_{\Ai}(\sigma + s, \sigma + y), \\
&& \!\!\!\!\!\!\!\!  \label{tildeT1} 
\tilde T_1(x,y) = \Delta^{1/3}  \int_{-\infty}^0  ds K_{\Ai}(\sigma_1 -s \Delta^{1/3} , \sigma_1 - x \Delta^{1/3}) K_{\Ai}(\sigma+s, \sigma + y) ,\\
&&  \!\!\!\!\!\!\!\!  \label{tildeS2} 
\tilde S_2(x,y) = K_{\Ai}(\sigma + x , \sigma + y) ,
\eea
where we have also changed $s \to \Delta^{1/3} s$ in the integrals. 
Since similarity transformations do not change the values of traces and determinants,
they will be useful below.

\section{Tail of Johansson's two-time distribution}

In this section we will expand Johansson's result \eqref{eq:KURTRESULT} for the JCDF 
for large positive values of $\sigma_1$ 
{\it at a given value of $\sigma$}.

\subsection{Limit $\sigma_1 \to +\infty$ of Johansson's formula}

We first consider the limit $\sigma_1 \to +\infty$ of the JCDF \eqref{eq:KURTRESULT}.
Since $\sigma$ is fixed it implies also $\sigma_2 \to +\infty$. Hence we should simply
obtain
\begin{equation}  \label{infinite} 
\lim_{\sigma_1 \to +\infty} \hat {\cal P}(\sigma_1,\sigma)  = 1,
\end{equation}
a simple check on the formula. 
Due to the presence of Airy kernels with $\sigma_1$ in their argument, in this limit we have  $S_1, T_1, S_3 \to 0$. Therefore neglecting sub-leading corrections in $\Ai(\sigma_1 + \ldots) \sim e^{-\frac{2}{3} \sigma_1^{3/2}}$ at large $\sigma_1$, we obtain  
\bea 
&& S(x,y) \to    \theta(x) S_2(x,y),  \\
&& T(x,y) \to \theta(x) S_2(x,y),  \\
&& R_u(x,y) \to\theta(x) S_2(x,y) ( 1    - u^{-1} ).
\eea
This implies that the Fredholm determinant in this limit is given by
\begin{align}
\text{Det} (1 + K_{u}) &  \to \text{Det} \left(
\begin{array}{cc}
 1 & 0  \\
 u P^+ R_u P^-  &  1+u P^+ R_u P^+ \\
\end{array}
\right) \nn \\&= \text{Det}   (1 - (1-u) P^{+} S_2 P^{+}  ) \nn 
\\&= \text{Det}   (1 - (1-u) P^{+} \tilde S_2 P^{+}  ) \nn
\\&= \text{Det}  (1 +  (u    - 1 ) P_{\sigma}K_{\Ai} P_{\sigma} ),
\end{align}
where we have used the similarity transformation \eqref{similarity} for $S_2$ and
the expression \eqref{tildeS2} for $\tilde S_2$, which is simply the Airy kernel.
To perform the integration over $u$ it is convenient to expand the Fredholm determinant in powers of its trace, namely 
\begin{equation}
\text{Det} (1 + f(u ) A) = 1+  \sum_{n=1}^\infty \frac{(f(u))^n}{n!} \left( \prod_{j=1}^n \int dx_j \right) \det_{i,j=1}^n A(x_i,x_j).
\end{equation} 
Therefore, pulling out the factor $ (u    - 1 )^n$ at each order, and using the identity  
\begin{equation}
\frac{1}{2 \pi i }\int_{R_1} \frac{d u}{ u-1}  (u-1 )^n = \delta_{n,0} \quad \forall n \geq 0,
\end{equation}
we obtain the expected result \eqref{infinite}.

\subsection{Large $\sigma_1$: leading tail contribution}

We now obtain the leading correction for large $\sigma_1$ at fixed $\sigma$. 
Let us first rewrite \eqref{defST} in the more compact form using the projectors
\bea \label{defST2} 
&& S = S_1  + P^+ S_2 - S_3 P^- \quad , \quad  T = -T_1 - P^+ S_2 + S_3 P^-,
\eea
which leads to {
\bea
\fl && \text{Det} (I + K_{u})   \\
\fl && = \text{Det} 
 \begin{pmatrix} I + P^- (S_1 - S_3 + u^{-1} (-T_1 + S_3) ) P^-   &  P^- (S_1 - u^{-1} T_1) P^+ \\
 P^+ (u (S_1 + S_2 - S_3) + (-T_1 - S_2 + S_3)) P^- &  I + P^+ (u(S_1 + S_2) + (-T_1 - S_2)) P^+
\end{pmatrix} .\nonumber 
\eea
}
In the large positive $\sigma_1$ limit, we can introduce a natural small parameter, namely $\epsilon \sim \Ai(\sigma_1)^2$. Then one has
\be
S_1 \sim T_1 \sim S_3 \sim \epsilon \quad , \quad S_2 \sim O(1).
\ee 
We will thus write the kernel as a sum of a $O(1)$ piece and a piece $\sim \epsilon$ as
\be
I + K_{u} =  \begin{pmatrix} I & 0 \\
C_0 & D_0 \end{pmatrix} +  \begin{pmatrix} A_1 & B_1 \\
C_1 & D_1 \end{pmatrix} = M_0 + M_1,
\ee
where we have defined
\bea
&& C_0 =  (u - 1) P^+ S_2 P^- \quad , \quad D_0= I + (u - 1) P^+ S_2 P^+,
\eea
and the perturbations which are $O(\epsilon)$
\bea
&& A_1  =  P^- \left( S_1 -  \frac{1}{u} T_1 +   S_3 \frac{1-u}{u} \right) P^- \\
&& B_1 = P^- \left(S_1 -  \frac{1}{u} T_1 \right) P^+ \\
&& C_1 = P^+ (u (S_1 - S_3) + (S_3-T_1)) P^- \\
&& D_1 =  P^+(u S_1 - T_1  )P^+ .
\eea
Using the expansion of the determinant around a fixed matrix
\be
\text{Det} (M_0 + M_1) = \text{Det} M_0 \left(1 + {\rm Tr} M_1 M_0^{-1} + \frac{1}{2} ( ({\rm Tr} M_1 M_0^{-1})^2 - {\rm Tr} (M_1 M_0^{-1})^2)
+ O(M_1^3)\right) ,
\ee
and the inversion formula $\begin{pmatrix} I & 0 \\
C_0 & D_0 \end{pmatrix}^{-1}=\begin{pmatrix} I & 0 \\
- D_0^{-1} C_0 & D_0^{-1} \end{pmatrix}$, we obtain to $O(\epsilon)$ \\
\be 
 \text{Det} (1 + K_{u}) = \text{Det} D_0 \, ( 1 + \text{Tr} D_1 D_0^{-1} +   \text{Tr} A_1 -   \text{Tr} B_1 D_0^{-1} C_0) +O(\epsilon^2), \label{expansion} 
\ee 
where the trace of an operator is defined as usual as
\begin{equation}
\text{Tr} A = \int dx A(x,x) = \int dy \tilde A(y,y)  \quad , \quad A(x,x')=\lambda \tilde A(\lambda x, \lambda x') ,
\end{equation}
and is invariant under the similarity transform $y=\lambda x$, a property used below.

We can now carry the integration over $u$ of \eqref{expansion} (since all operators there are $u$-dependent)
as in \eqref{eq:KURTRESULT}. It turns out that one only needs the formula
\be \label{int1} 
\frac{1}{2 \pi i }\int_{R_1} \frac{d u}{ u-1} \left( p_0(u) + \frac{1}{u} p_1(u) \right) = p_0(0) + p_1(0) - p_1(1) ,
\ee
where $p_0,p_1$ are arbitrary analytic functions in $u$. Indeed $A_1,B_1,D_1,C_0$ are of this form
with $p_0,p_1$ simple polynomials. Note that possible poles in the inverse $D_0^{-1}=(I + (u - 1) P^+ S_2 P^+)^{-1}$ would be canceled by the determinant $\text{Det} D_0$ which always appear in
factor. Hence we can expand $\text{Det} D_0$ and $D_0^{-1}$ to powers of $(u-1)$
and formally treat them as polynomials.

Using \eqref{int1} to integrate over $u$, we obtain
from \eqref{expansion} and \eqref{eq:KURTRESULT} for the various terms
\begin{align}
&\frac{1}{2 \pi i }\int_{R_1} \frac{d u}{ u-1} \text{Det} D_0 \,  \text{Tr} D_1 D_0^{-1} &  =  & \text{Tr} P^+ (S_1-T_1) P^+, \\
&\frac{1}{2 \pi i }\int_{R_1} \frac{d u}{ u-1}  \text{Det} D_0 \, \text{Tr} A_1 & = & \text{Tr}  P^- (S_1-T_1) P^- - F_2(\sigma) 
 \text{Tr}  P^- (S_3-T_1) P^- \nn,. \\
& \frac{1}{2 \pi i }\int_{R_1} \frac{d u}{ u-1}\text{Det} D_0 \,  \text{Tr} B_1 D_0^{-1} C_0 & = &
-F_2(\sigma) \text{Tr} P^- T_1 P^+ [1- P^+ S_2P^+]^{-1} P^+ S_2 P^- , \nn
\end{align}
where we used that $\text{Det} (1- P^+ S_2 P^+ )= F_2(\sigma)$. Putting everything together and simplifying we
finally obtain 
\bea  \label{eq:Johansonnlargesigma1}
 {\cal P}_\Delta(\sigma_1,\sigma)        &&
= 1 + \text{Tr} ( S_1 - T_1  )   + F_2(\sigma) \, \text{Tr} P^- (T_1-S_3),  \\
&&  + F_2(\sigma)\text{Tr} P^- T_1 P^+ (1- P^+ S_2 P^+)^{-1}  P^+ S_2   \quad +  O(\epsilon^2) \nn.
\eea

\section{Exact agreement between the two distributions in the large $\sigma_1$ limit}

\subsection{Summary}

Here we show that the two tails exactly agree. Let us recall our previous result for the tail of equation \eqref{eq:OURTAIL}.
Using that $(I-P_{\sigma} K_{\Ai}P_{\sigma} )^{-1} = I + (I-P_{\sigma} K_{\Ai}P_{\sigma} )^{-1} P_{\sigma}K_{\Ai} P_{\sigma}$ it can be rewritten as 
\begin{align} \label{eq:uslargesigma1}
\mathcal{P}_\Delta^{(1)} (\sigma_1,\sigma)  & =  1 +(F_2^{(1)}(\sigma_2)-1)    -   {F_2(\sigma)} \text{Tr} P_{\sigma_1} K_{\Ai} +  {F_2(\sigma)} \Delta^{1/3} \,  \text{\text{Tr}} P_\sigma K^{\Delta}_{\sigma_1}  \nn \\& +  {F_2(\sigma)}  \Delta^{1/3} \, \text{\text{Tr}} P_\sigma K_{\Ai} P_\sigma K^{\Delta}_{\sigma_1}   P_\sigma   (I-P_{\sigma} K_{\Ai}P_{\sigma} )^{-1}  .
\end{align}

We now compare Johansson's equation \eqref{eq:Johansonnlargesigma1} with our previous result   \eqref{eq:uslargesigma1}. As we show below one has
\begin{equation}\label{eq:rel1}
\text{Tr} ( S_1 - T_1  )  =  F_2^{(1)}(\sigma_2)-1,
\end{equation}
\begin{equation}\label{eq:rel2}
- \text{Tr} P^- S_3 = -   \text{Tr} P_{\sigma_1} K_{\Ai},
\end{equation}
\begin{equation}\label{eq:rel3}
\text{Tr} P^- T_1 = \Delta^{1/3}  \,  \text{Tr} P_\sigma K^{\Delta}_{\sigma_1} ,
\end{equation}
\begin{equation}\label{eq:rel4}
 \text{\text{Tr}} \,  P^- T_1 P^+ (1- P^+ S_2 P^+)^{-1}  P^+ S_2
=    \Delta^{1/3} \,  \text{\text{Tr}}  \, P_\sigma K_{\Ai} P_\sigma K^{\Delta}_{\sigma_1}   P_\sigma   (I-P_{\sigma} K_{\Ai}P_{\sigma} )^{-1}  .
\end{equation}

Hence the two formula \eqref{eq:OURTAIL} and \eqref{eq:Johansonnlargesigma1}
are identical. The tail obtained that we obtained in Ref. \cite{two-time-long-us,two-time-letter-us} 
is thus confirmed by the exact result of Johansson \cite{Johansson2times-final}.

\subsection{Proof of the above relations} \label{sec:proofs} 

We now show the above relations. We recall the definition \eqref{K4def0} 
of the kernel $K^{\Delta}_{\sigma_1}$. Since it involves $\Delta^{1/3}$ 
rather that $\Delta^{-1/3}$, it is more convenient to use the similarity transformed
functions $\tilde S_1,\tilde S_2,\tilde T_1$ defined in \eqref{STsim}
and use the invariance of the trace under the similarity transformation 
\eqref{similarity}. 

\subsubsection*{Proof of relation \eqref{eq:rel1}.}

For notational simplicity we use here and below notations such as 
$\int_{s} \equiv \int_{-\infty}^{+\infty} ds$, $\int_{s>0} \equiv \int_{0}^{+\infty} ds$,
$\int_{s,x>0} \equiv \int_{0}^{+\infty} ds \int_{0}^{+\infty} dx$, 
$\int_{s>0,x} \equiv \int_{0}^{+\infty} ds \int_{-\infty}^{+\infty} dx$
etc. One has 
\bea
\fl && \text{Tr}(S_1-T_1)  = \text{Tr}(\tilde S_1-\tilde T_1)  = - \int_{s,x} K_{\Ai}(\sigma_1 -s \Delta^{1/3} , \sigma_1 - x \Delta^{1/3}) K_{\Ai}(\sigma + s, \sigma + x) \nn \\
\fl && = - \int_{z_1,z_2>0,s,x} \Ai(\sigma_1 -s \Delta^{1/3}+z_1) \Ai(\sigma_1 - x \Delta^{1/3} + z_1)
\Ai(\sigma + s + z_2) \Ai(\sigma + x + z_2) .\nn
\eea
Now we use twice the identity
\bea
\int_s \Ai(a-s \Delta^{1/3}) \Ai(b+s) = \frac{1}{(1+\Delta)^{1/3}} \Ai\left(\frac{a + \Delta^{1/3} b}{(1+\Delta)^{1/3}}\right),
\eea 
and we obtain
\bea
\!\!\!\!\!\!\!\!\!\!\!\!\!\!\!\!\!\!\!\! \text{Tr}(S_1-T_1) & =& - \frac{1}{(1+\Delta)^{2/3}} 
\int_{z_1,z_2>0} 
\left[ \Ai\left(\frac{\sigma_1+z_1 + \Delta^{1/3} (\sigma+z_2)}{(1+\Delta)^{1/3}} \right) \right]^2 ,\nn \\
\!\!\!\!\!\!\!\!\!\!\!\!\!\!\!\!\!\!\!\!  & =& - \int_{z_1,z_2>0}  \left[ \Ai(z_1 + z_2 + \sigma_2)\right]^2
= -  \text{Tr} P_{\sigma_2} K_{\Ai} = F_2^{(1)}(\sigma_2)-1  ,
\eea
where $\sigma_2= \frac{\sigma_1 + \Delta^{1/3} \sigma}{(1+\Delta)^{1/3}}$ and we have changed
$z_{1,2} \to (1+\Delta)^{1/3} z_{1,2}$ in the integrals. 

\subsubsection*{Proof of relation \eqref{eq:rel2}.}
By definition from \eqref{defS3}.

\subsubsection*{Proof of relation \eqref{eq:rel3}.}

From \eqref{tildeT1} one has
\begin{align}
\text{Tr} P^- T_1  & = \text{Tr} P^- \tilde T_1 
= \Delta^{1/3}  \int_{x,s>0} K_{\Ai}(\sigma_1 +\Delta^{1/3} s , \sigma_1 + \Delta^{1/3} x) K_{\Ai}(\sigma - s  , \sigma - x  ),
\end{align}
where we have changed $(x,s) \to (-x,-s)$.
On the other hand, from the definition \eqref{K4def0}, our result is 
\begin{align}
 \Delta^{1/3} \text{\text{Tr}} P_\sigma K^{\Delta}_{\sigma_1} & =   \Delta^{1/3}
\int_0^\infty dv  \int_0^\infty dy_1 dy_2 \Ai\left(-  {y_1}{ }   +  v  + \sigma \right)  \\& \times  K_{\Ai}(y_1 \Delta^{1/3} + \sigma_1, y_2\Delta^{1/3} + \sigma_1)  \Ai \left(-  {y_2}{}  +  v  + \sigma\right) \nn \\& 
=   \Delta^{1/3} \int_0^\infty dy_1 dy_2 K_{\Ai}(y_1 \Delta^{1/3} + \sigma_1, y_2\Delta^{1/3} + \sigma_1)  
K_{\Ai}(-  {y_1}{ }     + \sigma, -  {y_2}{ }     + \sigma), \nn
\end{align}
and therefore they are the same. 

\subsubsection*{Proof of relation \eqref{eq:rel4}.}

We will prove the equivalent relation
\begin{equation}
 \text{\text{Tr}} \,  P^+ \tilde S_2 P^- \tilde T_1 P^+ (1- P^+ \tilde S_2 P^+)^{-1}  
=    \Delta^{1/3} \,  \text{\text{Tr}}  \, P_\sigma K_{\Ai} P_\sigma K^{\Delta}_{\sigma_1}   P_\sigma   (I-P_{\sigma} K_{\Ai}P_{\sigma} )^{-1}  .
\end{equation}
This calculation is not trivial due to the presence of the inverse. Given a generic operator $A$,  we use the definition of its inverse
\begin{equation}
(I-A)^{-1} = \sum_{n=0}^\infty A^n =  I + A + A^2 + \ldots.
\end{equation}
It is thus sufficient that for any $n \geq 0$
\begin{equation}\label{eq:reln}
 \text{\text{Tr}} \,  P^+ \tilde S_2 P^- \tilde T_1  (P^+ \tilde S_2)^n 
=    \Delta^{1/3} \,  \text{\text{Tr}}  \, P_\sigma K_{\Ai} P_\sigma K^{\Delta}_{\sigma_1}    (P_{\sigma} K_{\Ai} )^n,
\end{equation}
which we now prove. We start by showing the basic identity
\be \label{basic}
\int_{y_1>0} \Ai(\sigma+z-y_1) \tilde T_1(-y_1,y) = \Delta^{1/3} \int_{p_1>0} K^{\Delta}_{\sigma_1}(\sigma+z,\sigma+p_1) \Ai(\sigma+p_1+y) .
\ee
From the definition \eqref{tildeT1} we have (setting $s=-y_2$)
\bea
\fl && \Delta^{-1/3}   \int_{y_1>0} \Ai(\sigma+z-y_1) \tilde T_1(-y_1,y)  \\
\fl && 
= \int_{y_1,y_2>0}  \Ai(\sigma+z-y_1)  K_{\Ai}(\sigma_1 + y_2 \Delta^{1/3} , \sigma_1 + y_1 \Delta^{1/3}) 
K_{\Ai}(\sigma-y_2, \sigma + y) \nn \\
\fl && = \int_{p_1,y_1,y_2>0}  \Ai(\sigma+z-y_1)  K_{\Ai}(\sigma_1 + y_2 \Delta^{1/3} , \sigma_1 + y_1 \Delta^{1/3})  
\Ai(\sigma-y_2+p_1) \Ai(\sigma + y + p_1) \nn \\
\fl && =  \int_{p_1>0} K^{\Delta}_{\sigma_1}(\sigma+z,\sigma+p_1) \Ai(\sigma+p_1+y) ,
\eea
using the definition \eqref{K4def0}, which shows \eqref{basic}. Multiplying the l.h.s of \eqref{basic} by $\Ai(\sigma+z+x)$ and integrating over $z>0$ we obtain a second useful identity
\bea
\fl && (\tilde S_2 P^- \tilde T_1)(x,y)=
\int_{y_1>0} K_{\Ai}(\sigma+x,\sigma-y_1) \tilde T_1(-y_1,y) \\
\fl && = \Delta^{1/3}   \int_{z>0,p_1>0} \Ai(\sigma+z+x)
 K^{\Delta}_{\sigma_1}(\sigma+z,\sigma+p_1) \Ai(\sigma+p_1+y) .
 \eea
This can be rewritten more compactly, using the ket-bra notation for vectors and operators (as in quantum mechanics, i.e. $\langle x|\phi \rangle = \phi(x)$, 
$\langle \psi | O | \phi \rangle=\int_{xy} \psi(x) O(x,y) \phi(y)$ etc..) as
\bea \label{property} 
 (\tilde S_2 P^- \tilde T_1)(x,y)=  \Delta^{1/3}  \langle \Ai_{\sigma+x} | P^+ K^{\Delta}_{\sigma_1,\sigma} P^+ |  \Ai_{\sigma+y} \rangle,
\eea
where here and below we denote
\be
\Ai_{\sigma}(u)=\Ai(\sigma+u) \quad , \quad K^{\Delta}_{\sigma_1,\sigma}(u,v)=K^{\Delta}_{\sigma_1}(\sigma+u,\sigma+v) .
\ee 

Let us show \eqref{eq:reln} for $n=0$. One has
\bea
\fl  \text{\text{Tr}} \,  P^+ \tilde S_2 P^- \tilde T_1 
&=& \int_{x>0} (\tilde S_2 P^- \tilde T_1)(x,x) = \Delta^{1/3} \int_{x>0} 
\langle \Ai_{\sigma+x} | P^+ K^{\Delta}_{\sigma_1,\sigma} P^+ |  \Ai_{\sigma+x} \rangle \\
\fl &  =& \Delta^{1/3} \text{\text{Tr}} P^+  K_{\Ai,\sigma} P^+ K^{\Delta}_{\sigma_1,\sigma}
= \Delta^{1/3} \,  \text{\text{Tr}}  \, P_\sigma K_{\Ai} P_\sigma K^{\Delta}_{\sigma_1} ,
\eea
where we have used the definition of the Airy kernel
\be \label{def1} 
 K_{\Ai,\sigma} = \int_{z>0}  |  \Ai_{\sigma+z} \rangle \langle \Ai_{\sigma+z} |
\, \Leftrightarrow \,
K_{\Ai,\sigma}(x,y)= \langle x | K_{\Ai,\sigma} | y \rangle = \int_{z>0} \Ai(\sigma+z+x) \Ai(\sigma+z+y) .
\ee
We now turn to arbitrary $n$. One has
\bea \label{trace2} 
 \text{\text{Tr}} \,  P^+ \tilde S_2 P^- \tilde T_1  (P^+ \tilde S_2)^n = 
 \int_{x,y>0} (\tilde S_2 P^- \tilde T_1)(x,y) (P^+ \tilde S_2)^n (y,x).
\eea 
We now use the second equivalent definition of the Airy kernel
\bea
\tilde S_2(x,y) = K_{\Ai,\sigma}(x,y) = \langle  \Ai_{\sigma+x} |P^+| \Ai_{\sigma+y} \rangle,
\eea
which leads to
\bea \label{power} 
\fl && (P^+ \tilde S_2)^n (y,x) \\
\fl && = P^+  \int_{z_1,z_2,\dots,z_{n-1}>0}  
\langle  \Ai_{\sigma+y} |P^+| \Ai_{\sigma+z_1} \rangle 
\langle  \Ai_{\sigma+z_1} |P^+| \Ai_{\sigma+z_2} \rangle 
\cdots \langle  \Ai_{\sigma+z_{n-1}} |P^+| \Ai_{\sigma+x} \rangle .\nn
\eea
Inserting \eqref{property} and \eqref{power} in \eqref{trace2} we obtain
\bea
\fl  \text{\text{Tr}} \,  P^+ \tilde S_2 P^- \tilde T_1  (P^+ \tilde S_2)^n   &=&
\Delta^{1/3}  
\int_{x,y,z_1,z_2,\dots,z_{n-1}>0}  
\langle \Ai_{\sigma+x} | P^+ K^{\Delta}_{\sigma_1,\sigma} P^+ |  \Ai_{\sigma+y} \rangle \nn \\
\fl & \times  &
\langle  \Ai_{\sigma+y} |P^+| \Ai_{\sigma+z_1} \rangle 
\langle  \Ai_{\sigma+z_1} |P^+| \Ai_{\sigma+z_2} \rangle 
\cdots \langle  \Ai_{\sigma+z_{n-1}} |P^+| \Ai_{\sigma+x} \rangle \nn \\
\fl & =&   \Delta^{1/3} \,  \text{\text{Tr}}  \, P^+ K_{\Ai,\sigma} P^+ K^{\Delta}_{\sigma_1,\sigma}    (P^+  K_{\Ai,\sigma} )^n  \nn \\
\fl &=&  \Delta^{1/3} \,  \text{\text{Tr}}  \, P_\sigma K_{\Ai} P_\sigma K^{\Delta}_{\sigma_1}    (P_{\sigma} K_{\Ai} )^n,
\eea
where to go from the first two lines to the third we have used repeatedly \eqref{def1} 
for each of the $n+1$ integrals. 

\section{The case of different endpoints $X\neq 0$}

We consider now the case where the height at time $t_2$ is measured at $x=X \neq 0$ 
while the height at $t_1$ is measured at $x=0$, as in \eqref{defH12}. 
One defines the rescaled position variable 
\footnote{In the large $\Delta$ limit, the proper variable becomes $\bar{X}= \hat{X} \Delta^{-1/3}$.
However we do not focus on this limit here.}
\begin{equation}
\hat{X} = \frac{X}{2 (t_1 \Delta)^{2/3} } = \frac{X}{2 (t_2-t_1)^{2/3} },
\end{equation}
and the definitions \eqref{defh} still apply except that the definition of the scaled height difference variable $h$ becomes
\bea \label{defhX}
H_2-H_1= (\Delta t_1)^{1/3} ( h - \hat X^2) = \frac{X^2}{4 \Delta t_1} + (\Delta t_1)^{1/3} h,
\eea
where we recall $\Delta t_1=t_2-t_1$. We again define $\hat{\cal P}_\Delta(\sigma_1,\sigma_2)$ by Eq. \eqref{defP}. As in \cite{two-time-long-us}, the JPDF of $h_1$ and $h$, $P_\Delta(\sigma_1,\sigma)$,
is still defined by \eqref{defPP}, hence $\sigma$ is associated to $h$. The relation between
$\sigma_2$ and $\sigma$ is thus changed and one has
\be \label{PPhat3}
{\cal P}_\Delta(\sigma_1,\sigma) = 
\hat {\cal P}_\Delta \Big(\sigma_1,\sigma_2=\frac{\sigma_1 + (\sigma - \hat X^2) \Delta^{1/3}}{(1+\Delta)^{1/3}}\Big),
\ee 
and \eqref{eqdiff} is still valid. 

The Johansson formula \eqref{eq:KURTRESULT} is still valid for $X \neq 0$ \cite{Johansson2times-final} but with generalized functions
\footnote{we set there $\eta_1=0$, $\Delta \eta=\hat X$ and $\Delta \xi=\sigma- \hat X^2$ there, and use $\delta=\alpha \Delta \eta + 0^+$.} which we denote by an index $\hat X$. One has
\begin{equation} \label{S3X} 
S^{\hat{X}}_3(x,y) = e^{(y-x) \hat{X}  \Delta^{-1/3}} K_{\Ai}(\sigma_1 - x, \sigma_1 - y ).
\end{equation}
We display the others in their similarity transformed form, defined as in \eqref{similarity},
where we also changed $s \to \Delta^{1/3} s$ in the integrals

\begin{align}
\tilde S^{\hat{X}}_1(x,y)  = - \Delta^{1/3}   \int_0^\infty  ds& e^{(s-x) \hat{X} } 
K_{\Ai}(\sigma_1 -s \Delta^{1/3} , \sigma_1 - x \Delta^{1/3}) K_{\Ai}(\sigma  + s , \sigma +  y ),
\end{align}
\begin{align}  \label{T1X} 
\tilde T^{\hat{X}}_1(x,y)  =  \Delta^{1/3}   \int_{-\infty}^0  ds& e^{(s-x) \hat{X} } 
K_{\Ai}(\sigma_1 -s \Delta^{1/3} , \sigma_1 - x \Delta^{1/3}) K_{\Ai}(\sigma  + s , \sigma +  y ),
\end{align}
\begin{equation}
\tilde S^{\hat{X}}_2(x,y) = \tilde S_2(x,y) =  K_{\Ai}(\sigma  + x , \sigma+ y).
\end{equation}

%\begin{align}
%S^{\hat{X}}_1(x,y)  = - \Delta^{-1/3}   \int_0^\infty  ds& e^{\hat{X} (s-x) \Delta^{-1/3}} K_{\Ai}(\sigma_1 -s , \sigma_1 - x) \nn \\& \times K_{\Ai}(\sigma  + \hat{X}^2 + s \Delta^{-1/3}, \sigma  + \hat{X}^2 +  y \Delta^{-1/3} )
%\end{align}
%\begin{align}
%T^{\hat{X}}_1(x,y)  =  \Delta^{-1/3}  \int_{-\infty}^0  ds & e^{\hat{X} (s-x) \Delta^{-1/3}} K_{\Ai}(\sigma_1 -s , \sigma_1 - x) \nn \\& \times K_{\Ai}(\sigma +  \hat{X}^2  + s \Delta^{-1/3} , \sigma + \hat{X}^2 + y \Delta^{-1/3} )
%\end{align}
%\begin{equation}
%S^{\hat{X}}_2(x,y) = \Delta^{-1/3}  K_{\Ai}(\sigma+ \hat{X}^2  + x \Delta^{-1/3} , \sigma+ \hat{X}^2 + y \Delta^{-1/3})
%\end{equation}
%\begin{equation}
%S^{\hat{X}}_3(x,y) = e^{\hat{X}(y-x) \Delta^{-1/3}} K_{\Ai}(\sigma_1 - x, \sigma_1 - y )
%\end{equation}

Hence we see that, while $S_2$ is unchanged, in $\tilde S^{\hat{X}}_1$ and
$\tilde T^{\hat{X}}_1$  the Airy kernel $K_{\Ai}(\sigma_1 -s \Delta^{1/3} , \sigma_1 - x \Delta^{1/3})$
has been multiplied by $e^{\hat{X} (s-x)}$. By following the
same steps as for $\hat X=0$, it is easy to show that the large positive $\sigma_1$ tail is given 
by its prediction in \cite{two-time-long-us} as
\begin{align} 
\hat{\cal P}(\sigma_1,\sigma_2) = \mathcal{P}_{\Delta,\hat{X}}^{(1)} (\sigma_1,\sigma)
+ O(e^{- \frac{8}{3} \sigma_1^{3/2}})  ,
\end{align}
with
\begin{align} 
\mathcal{P}_{\Delta,\hat{X}}^{(1)} (\sigma_1,\sigma)   =  1 + & (F_2^{(1)}(\sigma_2')-1)    -   
{F_2(\sigma)} \text{Tr} P_{\sigma_1}K_{\Ai} +  {F_2(\sigma)}  \Delta^{1/3} \, \text{\text{Tr}} P_\sigma K^{\Delta, \hat{X}}_{\sigma_1}  \nn \\& +  {F_2(\sigma)} \Delta^{1/3} \, \text{\text{Tr}}  P_\sigma K_{\Ai} P_\sigma K^{\Delta, \hat{X}}_{\sigma_1}   P_\sigma   (I-P_{\sigma} K_{\Ai}P_{\sigma} )^{-1} ,\label{resX} 
\end{align}
i.e. as in \eqref{eq:uslargesigma1} where the kernel $K^{\Delta}_{\sigma_1}$ is replaced by
the $\hat{X}$-generalized kernel
 \begin{equation} \label{K4defX}
K^{\Delta,\hat{X}}_{\sigma_1}(u,v)   =  \int_0^\infty dy_1 dy_2 \Ai\left(-  {y_1}{ }   +  u \right) e^{\hat{X}(y_2- y_1)} K_{\Ai}(y_1 \Delta^{1/3} + \sigma_1, y_2\Delta^{1/3} + \sigma_1)  \Ai \left(-  {y_2}{}  +  v \right) ,
\end{equation}
exactly as defined in Eq. (140) in \cite{two-time-long-us}. In Eq. \eqref{resX} 
\be
\sigma_2' = \sigma_2 + \frac{ \Delta^{4/3}}{(1+\Delta)^{4/3}} \hat X^2 = \frac{H_2}{2 t_2^{2/3}} + \frac{X^2}{4 t_2^{4/3}} ,
\ee
as expected from the statistical tilt symmetry. Some details
of this derivation are reported in \ref{app:details}.

\section{Conclusion} 

In this paper we showed that the tail of the two-time joint distribution of the height for the KPZ universality class obtained in \cite{two-time-long-us,two-time-letter-us} is exact, as it agrees with the large $\sigma_1$ limit of the recently obtained rigorous two-time distribution \cite{Johansson2times-final}. This constitutes the first proof of the validity of the tail expression found in \cite{two-time-long-us,two-time-letter-us} for the droplet initial condition, which up to now was only checked indirectly by comparison with experimental and numerical data in \cite{two-time-letter-us}. The approximations in \cite{two-time-long-us,two-time-letter-us} consisted in performing, within the RBA method, a partial (instead of full) summation over the eigenstates of the LL model (keeping only the ground state for the first time slice from $t=0$ to $t=t_1$). We conjectured in \cite{two-time-long-us,two-time-letter-us}, based on previous experience on the one-time observables, that this gives the exact tail for large $\sigma_1$. Thanks to the exact result found in  \cite{Johansson2times-final}, we showed here that this conjecture is indeed correct. The expansion at large $\sigma_1$ started in this paper opens the way to obtain a full solution,
within the RBA method, of the multi-time problem, to be addressed in future publications.

\section*{Acknowledgment}
We are very grateful to K. Johansson for very helpful discussions and exchanges. \\
This work is supported by LabEx ENS-ICFP:ANR-10-LABX-0010/ANR-10-IDEX-0001-02 PSL* (J.D.N.) and  ANR grant ANR-17-CE30-0027-01 RaMaTraF (P.L.D.)

\bigskip

\appendix

\section{How to recover the JPDF of $h_1$ and $h$ from the cumulative distribution}
\label{app:jpdf} 
The JPDF of $h_1$ and $h$ can be obtained from the cumulative distribution {  $\hat {\cal P}_\Delta$} as follows 
(where the limit  $\lim_{t_1,t_2 \to +\infty,\frac{t_1-t_2}{t_1} =\Delta}$ is implicit) 
{ 
\bea
\fl && P_\Delta(\sigma_1,\sigma) = \overline{ \delta(h_1-\sigma_1)
\delta(h-\sigma) } = \frac{1}{(1+ \frac{1}{\Delta})^{1/3}} 
 \overline{ \delta(h_1-\sigma_1)
\delta(h_2- \frac{\sigma + \sigma_1 \Delta^{-1/3}}{(1+ \frac{1}{\Delta})^{1/3}} ) }  \\
\fl && = \frac{1}{(1+ \frac{1}{\Delta})^{1/3}}  \partial_{\sigma_1} \partial_{\sigma_2} \hat {\cal P}_\Delta(\sigma_1,\sigma_2) =
 \frac{1}{(1+ \frac{1}{\Delta})^{1/3}}  \partial_{\sigma_1} \partial_{\sigma_2} {\cal P}_\Delta(\sigma_1,\sigma=\sigma_2 (1 + \frac{1}{\Delta})^{1/3} - \sigma_1 \Delta^{-1/3}) \nonumber \\
\fl && = ( \partial_{\sigma_1} \partial_{\sigma} - \Delta^{-1/3} \partial_{\sigma}^2)  
{\cal P}(\sigma_1,\sigma) .\nonumber 
\eea
}
Note that $P_\Delta(\sigma_1,\sigma) =  \frac{1}{(1+ \frac{1}{\Delta})^{1/3}}  \hat P_\Delta(\sigma_1,\sigma_2)$,
where $\hat P_\Delta$ is the JPDF of $h_1$ and $h_2$
\bea
\hat P_\Delta(\sigma_1,\sigma_2) = \lim_{t_1,t_2 \to +\infty,\frac{t_1-t_2}{t_1} =\Delta} \overline{ \delta(h_1-\sigma_1)
\delta(h_2-\sigma_2) }.
\eea

\section{ {More details for the case} with non-zero endpoint $\hat{X} \neq 0$} \label{app:details} 

The large $\sigma_1$ expansion proceeds exactly as for $X=0$. Since 
$\tilde S_2$ is unchanged, it thus leads, as in \eqref{eq:Johansonnlargesigma1},
to the expression
\bea  \label{eq:Johansonnlargesigma1X}
 {\cal P}_\Delta(\sigma_1,\sigma)        &&
= 1 + \text{Tr} ( \tilde S_1^{\hat X} - \tilde T_1^{\hat X}   )   + F_2(\sigma) \, \text{Tr} P^- \tilde T_1^{\hat X} 
- F_2(\sigma) \, \text{Tr} P^- S_3^{\hat X} \\
&&  + F_2(\sigma)\text{Tr} P^- \tilde T_1^{\hat X} P^+ (1- P^+ \tilde S_2 P^+)^{-1}  P^+ \tilde S_2   \quad +  O(\epsilon^2) .\nn
\eea
Let us briefly examine each term. One has 
\bea
\fl &&  \text{Tr} ( \tilde S_1^{\hat X} - \tilde T_1^{\hat X}   )   \\
\fl && = - \int_{z_1,z_2>0,s,x} e^{(s-x) \hat X} \Ai(\sigma_1 -s \Delta^{1/3}+z_1) \Ai(\sigma_1 - x \Delta^{1/3} + z_1)
\Ai(\sigma + s + z_2) \Ai(\sigma + x + z_2) .\nn
\eea
We can now use the identity
\bea
&& \int_s e^{s \hat X} \Ai(a-s \Delta^{1/3}) \Ai(b+s) 
\times \int_x e^{-x \hat X} \Ai(a-x \Delta^{1/3}) \Ai(b+x)  \nn \\
&& = \frac{1}{(1+\Delta)^{2/3}} 
\left[ \Ai\left(\frac{a + \Delta^{1/3} b}{(1+\Delta)^{1/3}} - \frac{ \Delta^{1/3}}{(1+\Delta)^{4/3}} \hat X^2 \right) \right]^2,
\eea 
with $a=\sigma_1+ z_1$ and $b=\sigma+z_2$. Proceeding as in Section \ref{sec:proofs} 
one obtains
\be
  \text{Tr}(\tilde S_1-\tilde T_1)  = -  \text{Tr} P_{\sigma_2'} K_{\Ai} = F_2^{(1)}(\sigma_2')-1  
\quad , \quad \sigma_2' = \sigma_2 + \frac{ \Delta^{4/3}}{(1+\Delta)^{4/3}} \hat X^2,
\ee
where we have used that 
$\sigma_2=\frac{\sigma_1 + (\sigma - \hat X^2) \Delta^{1/3}}{(1+\Delta)^{1/3}}$. \\

Moreover we find, since the prefactor in \eqref{S3X} does not affect the trace of the kernel
\begin{equation}
\text{Tr} P^- S_3^{\hat X} =   \text{Tr} P_{\sigma_1} K_{\Ai}
\end{equation}
Next, from \eqref{T1X} one has
\begin{align} \label{top}
\text{Tr} P^- \tilde T_1^{\hat X} 
= \Delta^{1/3}  \int_{x,s>0} e^{(x-s) \hat X} K_{\Ai}(\sigma_1 +\Delta^{1/3} s , \sigma_1 + \Delta^{1/3} x) K_{\Ai}(\sigma - s  , \sigma - x  ).
\end{align}
On the other hand, from the definition \eqref{K4defX}, our result is 
\begin{align}
 \Delta^{1/3} \text{\text{Tr}} P_\sigma K^{\Delta,\hat X}_{\sigma_1} & =   \Delta^{1/3}
\int_0^\infty dv  \int_0^\infty dy_1 dy_2 \Ai\left(-  {y_1}{ }   +  v  + \sigma \right)  \\& \times  
e^{\hat X(y_2-y_1)} K_{\Ai}(y_1 \Delta^{1/3} + \sigma_1, y_2\Delta^{1/3} + \sigma_1)  \Ai \left(-  {y_2}{}  +  v  + \sigma\right) \nn \\& 
=   \Delta^{1/3} \int_0^\infty dy_1 dy_2 e^{\hat X(y_2-y_1)} K_{\Ai}(y_1 \Delta^{1/3} + \sigma_1, y_2\Delta^{1/3} + \sigma_1)  \nn \\ \times &
K_{\Ai}(-  {y_1}{ }     + \sigma, -  {y_2}{ }     + \sigma), \nn
\end{align}
which is identical to \eqref{top}.

Finally, as in Section \ref{sec:proofs}, one can prove 
\begin{equation}\label{eq:reln2}
 \text{\text{Tr}} \,  P^+ \tilde S_2 P^- \tilde T_1^{\hat X}  (P^+ \tilde S_2)^n 
=    \Delta^{1/3} \,  \text{\text{Tr}}  \, P_\sigma K_{\Ai} P_\sigma K^{\Delta,\hat X}_{\sigma_1}    (P_{\sigma} K_{\Ai} )^n.
\end{equation}
It is easy to see, following the same steps, that \eqref{property} becomes 
\bea \label{property2} 
 (\tilde S_2 P^- \tilde T_1^{\hat X})(x,y)=  \Delta^{1/3}  \langle \Ai_{\sigma+x} | P^+ K^{\Delta,\hat X}_{\sigma_1,\sigma} P^+ |  \Ai_{\sigma+y} \rangle.
\eea
The same steps as in Section \ref{sec:proofs} then immediately lead to \eqref{eq:reln2}.
Hence we obtain that \eqref{eq:Johansonnlargesigma1X} is identical to our result
\eqref{resX}-\eqref{K4defX}.

\section*{References}

\bibliographystyle{iopart-num}
\bibliography{two_times_references}

\end{document}